\documentclass[floatfix,pra,amsmath,amssymb,letterpaper,groupaddresses,superscriptaddress,twocolumn]{revtex4}
\usepackage{times}
\usepackage{latexsym}
\usepackage{graphicx}
\usepackage{amsmath,amssymb}
\usepackage{verbatim,times,bbm}
\usepackage{soul}
\usepackage[colorlinks,hyperindex]{hyperref}
\hypersetup{colorlinks,citecolor=blue,linkcolor=blue,urlcolor=blue}

\usepackage{color}

\usepackage{epsf,epsfig}
\usepackage{ifpdf}
\usepackage[utf8]{inputenc}
\usepackage[T1]{fontenc}
\usepackage{graphicx}
\input{epsf}
\usepackage{epsfig}
\usepackage{epstopdf}
\usepackage{arcs}
\usepackage{subfigure}
\usepackage{tensor}
\usepackage{braket}
\usepackage{float}
\usepackage[dvipsnames]{xcolor}
\usepackage{amsmath}
\usepackage{morefloats}
\newcommand{\be}{\begin{equation}}
	\newcommand{\ee}{\end{equation}}

\usepackage{tikz}
\usetikzlibrary{calc}   
\usetikzlibrary{topaths}

\usepackage{hyperref}
\hypersetup{
	colorlinks =WildStrawberry,
	linkcolor =red,
	pdftitle={Leading corrections to Kerr geometry},
	citecolor=NavyBlue,
	filecolor=WildStrawberry,
	urlcolor=WildStrawberry,
}

\usepackage[toc]{appendix}
\usepackage{color,soul}
\usepackage{datetime}

\newcommand{\bea}{\begin{eqnarray}}
	\newcommand{\eea}{\end{eqnarray}}
\newcommand{\ba}{\begin{eqnarray}}
	\newcommand{\ea}{\end{eqnarray}}

\newcommand{\beq}{\begin{equation}}
	\newcommand{\eeq}{\end{equation}}
\newcommand{\beqa}{\begin{eqnarray}}
	\newcommand{\eeqa}{\end{eqnarray}}
\newcommand{\beqar}{\begin{eqnarray*}}
	\newcommand{\eeqar}{\end{eqnarray*}}

\renewcommand{\href}[2]{#2}

\begin{document}

\title{Irreversible Entropy Production rate  in a parametrically driven-dissipative System: The Role of Self-Correlation between Noncommuting Observables}

\author{Sareh Shahidani\footnote{sareh.shahidani@gmail.com}}
\affiliation{Department of Physics, Sharif University of Technology, Tehran 14588, Iran}
\author{Morteza Rafiee\footnote{m.rafiee@shahroodut.ac.ir}}
\affiliation{Faculty of Physics, Shahrood University of Technology , 3619995161 Shahrood, Iran}

\vspace{0.1cm}

\begin{abstract}
	In this paper, we explore the Wigner entropy production rate in the stationary state of a two-mode Gaussian system. The interacting modes dissipate into different local thermal baths. Also, one of the bosonic modes evolves into the squeezed-thermal state because of the parametric amplification process.Using the Heisenberg-Langevin approach, combined with quantum phase space formulation, we get an analytical expression for the steady-state Wigner entropy production rate. It contains two key terms. The first one is an Onsager-like expression that describes heat flow within the system. The second term resulted from vacuum fluctuations of the baths. Analyses show that self-correlation between the quadratures of the parametrically amplified mode pushes the mode towards the thermal squeezed state. It increases vacuum entropy production of the total system and reduces the heat current between the modes. The results imply that, unlike other previous proposals, squeezing can constrain the efficiency of actual non-equilibrium heat engines by irreversible flows.
\end{abstract}

\maketitle

\section{Introduction}
Entropy production rate, which is related to the second law of thermodynamics  plays a fundamental role in non-equilibrium classical and quantum thermodynamics. It provides a good framework  to describe and quantify irreversibility of physical phenomena. For a system in contact with a bath, the variation of entropy per unit time can be expressed as 
\begin{equation}
\frac{dS(t)}{dt}=\Pi(t)-\Phi(t),
\end{equation}
where $S$ is the entropy of the system, $\Phi(t)$ is the flow of entropy between the system and its bath (per unit time) and  $\Pi(t)$ is  the entropy production rate.  When a system reaches a non-equilibrium steady-state $dS(t)/dt$ vanishes,  and we have $\Pi_s=\Phi_s> 0$. If the stationary state is a thermal equilibrium state, then $\Pi_s=\Phi_s =0$. 

So far, the theory of entropy production has been developed and formulated in different contexts. In the classical domain, 
 the most popular approaches are based on  Onsager's theory\cite{Onsager, Machlup,Tisza}, classical master equations \cite{Schnakenberg, Tome1}, and Fokker-Planck equations\cite{Van den Broeck,Tome2,Spinney,Seifert,Landi}.
The extension of these approaches to the quantum domain leads to new formulations  based on the quantum master equations \cite{Gorini,Lindblad,Spohn,Leggio}, quantum trajectories \cite{Elouard}, and fluctuation theorems \cite{Jarzynski, Crooks}. 

More recently, considerable attention has been directed towards the theoretical and experimental characterization of entropy production in non-equilibrium bosonic systems  based on the quantum phase space distributions and Fokker-Planck equations\cite{Santos1, Brunelli1, Santos2,Santos3,Malouf,Zicari,Giordano,Brunelli2,Malouf-inf,Salazar,Rossi}.

 In this framework, one can express entropy production in terms of means and variances of independent variables and identify  irreversible quasi-probability currents in phase space for single-mode Gaussian systems in contact  with a single reservoir\cite{Santos1, Santos3} and multi-mode  Gaussian systems connected to multiple reservoirs\cite{Brunelli1,Malouf,Zicari,Giordano}. Identifying the  quasi-probability currents  in phase space provides physical interpretations for the irreversibility in the microscopic systems  at the quantum level\cite{Santos3}. It can keep   Wigner entropy production rate finite for a system in contact with a thermal reservoir at zero temperature\cite{Santos2,Malouf}.
 Also, the formalism works very well to describe irreversibility in  open quantum systems exposed to different non-equilibrium reservoirs, such as squeezed and dephasing reservoirs\cite{Santos1}. 
 
This paper is motivated by the above studies and recent proposals \cite{Huang,Rosnagel,Abah,Alicki,Niedenzu1,Klaers,Agarwalla,Niedenzu2,Manzano-16,Manzano-sq,Manzano-2021} of using squeezed thermal  states as a resource to increase the efficiency of heat engines beyond the Carnot limit.  
 A squeezed-thermal reservoir can be obtained by applying a unitary transformation $U_{sq}$ to a thermal reservoir $\rho_{th-sq}=U_{sq}\rho_{th}U_{sq}^{\dagger}$. However, any squeezing operation increases the energy of the thermal bath. It is believed that  the second law of thermodynamics remains valid in these machines if the required energy is accounted for.

 On the other hand, obtaining Carnot efficiency and surpassing it has  limited significance for practical applications. Since the reversible process has to be infinitely slow and the corresponding power is zero. In practical applications, when the cycling process takes place in a finite time, one should recognize the sources of irreversibility within the system. The system considered in this study is a step towards understanding the limits that thermodynamics imposes on the efficiency of  squeezed-thermal  machines for practical applications in the quantum regime. 

 The system under study is composed of two harmonic oscillators which are linearly coupled and dissipate into different local thermal baths. One oscillator is parametrically driven.  Hence, the degenerate parametric amplification process  which is responsible for generating squeezed-thermal states, happens within the system. By applying a squeezing transformation to this system, one can diagonalize the Hamiltonian and obtain a unitarily equivalent system comprising  two linearly coupled harmonic oscillators dissipating into a thermal and a squeezed-thermal bath. Identifying the sources of irreversibility for each of  these systems and comparing them are the steps toward determining the required energy for constructing non-equilibrium squeezed bath and calculating the practical efficiencies of the non-equilibrium squeezed-thermal machines. 
    
 Using the quantum phase space method, we obtain an analytical expression for the Wigner entropy production of the system. 
  
It has been shown that, in  composite quantum systems, the entropy production is related to the internal correlation between subsystems \cite{Brunelli2,Malouf, Malouf-inf}. We show that the parametric amplification process not only modifies the internal correlation between the subsystems but also produces an additional correlation between non-commuting observables ($q$ and $p$) of the parametrically driven oscillator. This self-correlation term, which is responsible for  squeezing of the mode, also contributes to the entropy production. To analyze the role of squeezing, we decompose the Wigner entropy production into two terms.  The first term is an Onsager-like expression that describes heat flow between the subsystems and  the second one describes the irreversibility due to vacuum fluctuations of both thermal baths. The second contribution is always nonzero, even when both reservoirs are in the vacuum state. 

We demonstrate that  the self-correlation between quadratures ($q$ and $p$) of the driven  oscillator produces a heat current from cold to hot bath reduces the effective heat current between them. Also, analyzing the vacuum part of the entropy production allows us to discriminate the local and non-local effects of the baths on the oscillators. To support our analytical results, we calculate vacuum entropy production and its components in two different regimes. In both regimes, self-correlation increases the local contribution of the vacuum entropy production, the internal correlation increases the non-local contribution of the reservoirs, and they compete with one another. 

The paper is structured as follows: In section \ref{Sec:Model}, by introducing the physical model, we derive the quantum Langevin equations of motion for the system operators. In section \ref{sec:CM}, we derive an analytical expression for steady-state Wigner entropy production in terms of self-correlation and internal correlation terms of the covariance matrix.
Then, we identify different sources of irreversibility within the system.
 In section \ref{NUMERIC}, we support our analytical results by numerical calculations to show the effect of squeezing on the Wigner entropy production in two different regimes.  Section \ref{Sec:Conc} contains the conclusions of the study.

\section{The Model}\label{Sec:Model}
\begin{figure}
\centering
 \includegraphics[width=8cm,height=2.4cm,angle=0]{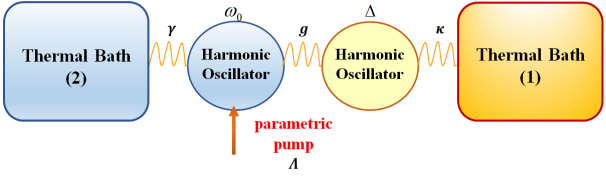}
 \caption{(Color online) Schematic of the system.}
 \label{Fig1-Scheme} 
\end{figure}

The  system consists  of two coupled  harmonic quantum oscillators. Each oscillator dissipates into its local thermal bath. The oscillations of one of the harmonic oscillators is parametrically amplified by an external force (Fig. \ref{Fig1-Scheme}). The unitary dynamics of the system can be described by the following Hamiltonian 
\begin{eqnarray}\label{H1}
H=\Delta \hat{a}^\dagger \hat{a} +\omega_0 \hat{b}^\dagger \hat{b} + \Lambda(\hat{b}^2+\hat{b}^{\dagger 2}) -g(\hat{a}+\hat{a}^\dagger)(\hat{b}+\hat{b}^\dagger),\nonumber\\
\end{eqnarray}
where  $a(a^{\dagger})$ and $b (b^{\dagger})$ are the annihilation (creation) operators of the harmonic oscillators with frequency $\Delta$ and  $\omega_0$, respectively. The first two terms of Eq. (\ref{H1}) describe the free energy of harmonic  oscillators, the third term describes the parametric amplification of the harmonic oscillator with frequency $\omega$ and strength $\Lambda$, and the last term describes the  coupling between the harmonic oscillators with coupling constant $g$.  The first harmonic oscillator  is locally connected to thermal bath 1 and dissipates at  rate   $\kappa$, and the second one is locally connected to  thermal bath 2 and dissipates at  rate  $\gamma$. 

In practice, the system can be an optomechanical system that contains a $\chi^{(2)}$ medium \cite{Agarwall} or an optomechanical system with Duffing nonlinear mechanical oscillator\cite{Lu}. In these systems, the  fluctuations of the operators of the subsystems (cavity-field and mechanical oscillator) around the mean-field values play the role of the Gaussian modes\cite{Brunelli1,Brunelli2,Rossi}. 

Considering the conventional picture of  system-bath  coupling in the form
\begin{equation}
H_{SB}=i\hbar \int_{-\infty}^{\infty} d\omega\kappa(\omega)(B^{\dagger}(\omega)c-c^{\dagger}B(\omega)),
\end{equation}
to describe the dissipation of each mode at the level of standard input-output theory \cite{Gardiner1, Gardiner-book}, we can get the damping of each mode.
We use the Heisenberg-Langevin approach to describe the dynamics of the system \cite{Gardiner-book}. In a recent paper\cite{Konopik}, an excellent agreement  has been found between the Langevin-Heisenberg approach and the global Lindblad master equation in such systems.
 According to this approach, including dissipation caused by system-bath couplings and the corresponding noises, quantum Langevin equations of motion, are given by
\begin{subequations}\label{Langevin_1}
\begin{eqnarray}
\dot{\hat{a}}&=&-(i \Delta + \kappa) \hat{a} + i g (\hat{b}+\hat{b}^\dagger)+ \sqrt{2\kappa} a_{in},\\
 \dot{\hat{b}}&=&-(i \omega_0+ \gamma)\hat{b} -2 i \Lambda \hat{b}^\dagger +i g (\hat{a} + \hat{a}^\dagger) +\sqrt{2\gamma} b_{in},\nonumber\\
\end{eqnarray}
\end{subequations}
  where  $a_{in}$ and $b_{in}$ are the input thermal noise operators characterized by the following Markovian correlation functions  
\begin{subequations}
 \begin{eqnarray}\label{correlation-th}
 < a_{in}(t)a_{in}(t')>&=& < a^{\dagger}_{in}(t)a^{\dagger}_{in}(t')> =0,\\
< a_{in}(t)a_{in}^{\dagger}(t')> &=&(\bar{n}_{1}+1)\delta(t-t'),\\
 < b_{in}(t)b_{in}(t')> &=&< b_{in}^{\dagger}(t)b^{\dagger}_{in}(t')>=0,\\
< b_{in}(t) b_{in}^{\dagger}(t')> &=&(\bar{n}_{2}+1)\delta(t-t'),
\end{eqnarray}
\end{subequations}
 where the thermal occupation number corresponding to reservoirs at temperature $T_i(i=1,2)$ are given by
\begin{equation}
\bar{n}_{1}=(e^{\hbar \Delta/k_B T_1}-1)^{-1},\bar{n}_{2}=(e^{\hbar \omega_0/k_B T_{2}}-1)^{-1}, 
\end{equation}
Defining $x=(\hat{a}+\hat{a}^\dagger)/\sqrt{2}$, $y=(\hat{a}-\hat{a}^\dagger)/i \sqrt{2}$, as the quadratures of the first mode,  $q=(\hat{b}+\hat{b}^\dagger)/\sqrt{2}$, $p=(\hat{b}-\hat{b}^\dagger)/i\sqrt{2}$ as the quadratures of the second mode (and similarly their corresponding input noise quadratures) and  coupling strength $G=2g$, quantum Langevin equations for the quadratures can be written in the compact matrix form  
\begin{equation}\label{eq_motion}
 \dot{u}(t)=Au(t) +f(t).
\end{equation}
where $u(t)=(x,y,q,p)^T$ is the vector of operators and $f(t)=(\sqrt{2\kappa}x_{in},\sqrt{2\kappa}y_{in},  \sqrt{2\gamma}q_{in}, \sqrt{2\gamma}p_{in})^T$  is the vector of noises and the drift matrix A is given by
\begin{equation}\label{drift}
A =\left(\begin{array}{*{20}c}
{{-\kappa}} & {{\Delta}} & {{0}} & {{0}}  \\
{{-\Delta}} & {{-\kappa}} & {{ G}} & {{0}}  \\
{{0}} & {{0}} & {{-\gamma}} & {{\omega}}  \\
{{ G}} & {{0}} & {{-\Omega}} & {{-\gamma}}  \\
\end{array}\right).
\end{equation}
where $\omega\equiv\omega_0-2\Lambda$ and $\Omega\equiv\omega_0+2\Lambda$.
The formal solution of  equation (\ref{eq_motion}) is $u(t)=M(t)u(0)+\int_0^tds M(s)f(t-s),$ where $M(t)=\exp\{At\}$. The system is stable and reaches a steady-state if all  eigenvalues of the drift matrix A have negative real parts.
For  $\Delta,\omega>0$, applying the Routh–Hurwitz criterion\cite{DeJesus}, gives the following stability condition:
\begin{equation}\label{stability-1}
\eta:=\delta^2 \Gamma^2-G^2\Delta\omega>0,
\end{equation}
where $\eta$ is the stability parameter, $\delta=\sqrt{\Delta^2+\kappa^2}$ and $\Gamma=\sqrt{\gamma^2+\omega\Omega}$ .
\section{Correlation Matrix and Wigner Entropy Production}\label{sec:CM}
Since the dynamics of the system is linear and the quantum noise terms are Gaussian, the steady-state of the  system is a continuous variable (CV) Gaussian state, which can be characterized by the $4\times 4$ correlation matrix (CM) $\sigma$ with corresponding components $\sigma_{i,j}= <u_i (\infty)u_j (\infty) +
u_j (\infty)u_i (\infty)>/2$ (with the assumption that the first moments are zero). When the system is stable, each component of the CM is given by 
\begin{equation}\label{Eq:sigmaij}
\sigma_{i,j}=\Sigma_{k,l}\int_0^{\infty}ds\int_0^{\infty}ds' M_{ik}(s)M_{jl}(s')\phi_{kl}(s-s'),
\end{equation} 
where  $\phi(s-s')$ is the matrix of the stationary noise correlation functions. It should be noted that  the uncertainty relations among canonical operators ($[u_i,u_j]=i\Omega_{ij}$) impose a constraint on the quantum  CMs, corresponding to the inequality $\sigma+i \bf{\Omega}/2\succeq0$, where $\bf{\Omega}$ is the symplectic matrix\cite{Serafini-BOOK}. 

When the stability condition is satisfied, $M(\infty)=0$ and one gets the  Lyapunov equation for the steady-state correlation matrix as 
\begin{equation}\label{lyap}
 A \sigma + \sigma A^T =- D,
\end{equation}
where $D=diag \{\kappa (2\bar{n}_{1}+1), \kappa (2\bar{n}_{1}+1), \gamma (2\bar{n}_{2}+1), \gamma (2\bar{n}_{2}+1) \}$  is the diffusion matrix.
Equation (\ref{lyap}) which  is linear in $\sigma$  is  exactly solvable in a straightforward manner, but the general exact expression is too cumbersome and will not be presented here. 

The Gaussian nature of the system allows us to relate  Wigner entropy of the system to the  covariance matrix  ($\sigma$) by
\begin{equation}\label{W-form}
w_{\sigma}(u)=\frac{1}{(2\pi)^n\sqrt{det\sigma}}e^{-\frac{1}{2}u^T\sigma^{-1}u},
\end{equation}
which is always positive  and allows us to identify  Wigner function as a quasi-probability distribution in phase space ($n$ is the number of bosonic modes). 
It provides a fully equivalent description of the density matrix. 
These features  make the   Wigner entropy, introduced  in Ref. \cite{Santos1}, as 
\begin{equation}\label{W-entropy}
S_w=-\int du w_{\sigma}(u)\log w_{\sigma}(u)
\end{equation}
a perfectly suitable framework for quantification of irreversibility in the current system. It is  shown in Ref. \cite{Adesso} that Wigner entropy  can be related to the Rényi-2 entropy and satisfies the strong subadditivity inequality. The link between general Rényi-$\alpha$ entropies and the thermodynamic properties of quantum systems has also been of interest in recent years\cite{Wei, Brandao}. 

The dynamics of the system can be equivalently expressed in terms of the Fokker-Planck equation for the Wigner function of Eq. (\ref{W-form}). Therefore, following the approach of Refs. \cite{Landi,Brunelli1,Brunelli2}, the entropy production rate ($\Pi_s$) and entropy flux rate ($\phi_s$) in the steady-state take the following form
\begin{eqnarray}\label{Pi-s-1}
\Pi_s=-\phi_s&=&Tr(2 A^{irr} D^{-1} A^{irr} \sigma^s + A^{irr})\nonumber\\
&=&2\kappa(\frac{\sigma_{11}^s+\sigma_{22}^s}{N_1}-1)+2\gamma(\frac{\sigma_{33}^s+\sigma_{44}^s}{N_2}-1),\,\,\,\,\,\,\,\,\,\,\,
\end{eqnarray}
where  $A^{irr}=diag\{-\kappa,-\kappa,-\gamma,-\gamma\}$ and $N_i=(1+2\bar{n}_i), (i=1,2)$. 
When the system is in the equilibrium state we have $\sigma_{11}^s+\sigma_{22}^s=N_1$ and $\sigma_{33}^s+\sigma_{44}^s=N_2$, and hence,  $\Pi_s=0$.
In this view, Eq.  (\ref{Pi-s-1}) can be expressed as $\Pi_s=\sum_{i=1,2}\Pi_i^s$  and  is a suitable quantifier of the distance between  the bath-imposed equilibrium state of each oscillator and  its  steady-state \cite{Brunelli2,Malouf, Salazar}. However, we shall show that $\Pi_s$ is not always capable of detecting self-correlation between quadratures of each subsystem.  

It should be noted that according to the Lyapunov equation (\ref{lyap}), in the steady-state the diagonal terms($\sigma_{ii}$) and off-diagonal ($\sigma_{ij\neq i}=\sigma_{ji\neq i}$) terms of the CM are not independent. In fact, it is easy to show that 
\begin{subequations}\label{vii}
\begin{eqnarray}
\sigma_{11}^s&=&\frac{N_{1}}{2}+\frac{\Delta}{\kappa}\sigma_{12}^s,\\
\sigma_{22}^s&=&\frac{N_{1}}{2}-\frac{\Delta}{\kappa}\sigma_{12}^s+\frac{G}{\kappa}\sigma_{23}^s,\\
\sigma_{33}^s&=&\frac{N_{2}}{2}+\frac{\omega}{\gamma}\sigma_{34}^s,\\
\sigma_{44}^s&=&\frac{N_{2}}{2}-\frac{\Omega}{\gamma}\sigma_{34}^s+\frac{G}{\gamma}\sigma_{14}^s,
\end{eqnarray}
\end{subequations}
and hence the internal correlation between the quadratures of the two modes ($\sigma_{14}^s$ and $\sigma_{23}^s$) and self-correlation between quadratures of each mode  ($\sigma_{34}^s$ and $\sigma_{12}^s$) are concealed in the full expression of the diagonal elements.
Besides, the correlation terms result in a steady-state with reduced uncertainty in one quadrature and enhanced uncertainty in the  complementary conjugated-quadrature for each mode. This implies that, in general, the steady-state of each bosonic mode is a thermal-squeezed state.

Hence, the entropy production can be related to off-diagonal elements of the CM as 
\begin{equation}\label{Pi_s2}
\Pi_s=\frac{2G}{N_2}\sigma_{14}^s+\frac{2G}{N_{1}}\sigma_{23}^s-\frac{2(\Omega-\omega)}{N_2}\sigma_{34}^s.
\end{equation}
Equation (\ref{Pi_s2}) quantitatively  links  the steady-state entropy production rate to the correlation functions of the dynamical variables. The first two terms of $\Pi_s$, which  are nonzero only for $G\neq0$, represent the effect of coupling between the two modes and the correlation between their quadratures and the last term, which  is nonzero only for $\Omega\neq\omega$, represents the effect of the self-correlation between the quadratures of the parametrically amplified mode. As can be seen, the self-correlation between the quadratures of the first mode $\sigma_{12}^s$  does not contribute to $\Pi_s$. Therefore, $\Pi_s$ is incapable of detecting self-correlation between the quadratures of both modes. 

It should be noted that all correlation terms in $\Pi_s$ are functions of the internal coupling constant ($G$), the strength of the parametric drive ($\Lambda$), the mean occupation number of reservoirs ($N_1, N_2$), and dissipation rates ($\kappa,\gamma$).  
The exact analytical expressions for these terms are too cumbersome and not tractable. Therefore,  we leave the details for the appendix. For the purpose of illustration, we present here the correlation terms ($\sigma_{ij}$) and steady-state entropy production in compact forms. The correlation terms can be compactly cast as 
\begin{subequations}\label{co-terms-1}
\begin{eqnarray}
\sigma_{14}^s&=&\sigma_{41}^s=(a_{11}N_1+a_{12}N_2)/2,\\
\sigma_{23}^s&=&\sigma_{32}^s=(a_{21}N_1+a_{22}N_2)/2,\\
\sigma_{34}^s&=&\sigma_{43}^s=(a_{31}N_1+a_{32}N_2)/2,
\end{eqnarray} 
\end{subequations}
Therefore, we can express $\Pi_s$ as
\begin{equation}\label{Pi-s-final}
\Pi_s=\Pi_0+\Pi_1,
\end{equation}
 where 
\begin{subequations}\label{Pi_0,1}
\begin{eqnarray}
\Pi_0&=&G(a_{12}+a_{11}+a_{22}+a_{21})-4\Lambda (a_{32}+a_{31}),\quad\quad\\
  \Pi_1&=&(\frac{G a_{11}-4\Lambda a_{31}}{N_2}-\frac{G a_{22}}{N_2})(N_1-N_2),
\end{eqnarray} 
\end{subequations}
Equations (\ref{Pi-s-final}) and  (\ref{Pi_0,1}) are the main results of this study.
In the decomposition of $\Pi_s$, $\Pi_0$ can be regarded as  the vacuum part of the entropy production, which is not dependent on the mean excitation number of the baths and is not zero even for zero temperature baths ($N_1=N_2=1$).

 For $N_1=N_2=1$, we can express entropy production as $\Pi_0=\sum\limits_{k=1}\limits^{3}j_k$ where 
 \begin{eqnarray}\label{jk-node-node}
 j_1&=&<\frac{d}{dt}a^{\dagger}a>=i<[H,a^{\dagger}a]>=2G\sigma_{14}^{s},\\
 j_2&=&<\frac{d}{dt}b^{\dagger}b>=i<[H,b^{\dagger}b]>\nonumber\\&=&2G\sigma_{23}^{s}-(\Omega-\omega)\sigma_{34}^s,\\
 j_3&=&-(\Omega-\omega)\sigma_{34}^s,
 \end{eqnarray}
In these expressions, $j_1$ and $j_2$ are known as excitation currents due to node-node coupling in the lattices \cite{Malouf}. In the current system, the counter-rotating terms of the interaction Hamiltonian $(a^{\dagger}b^{\dagger}+ H.c.)$, do not conserve the number of excitations in the system. Hence, the unitary internal dynamics maintains a stationary current within the system, that is non-zero for $N_1=N_2=1$ and $\Lambda=0$. In other words, these two currents  are related to intersystem correlations $\sigma_{14}$ and $\sigma_{23}$ that can be quantified by mutual information. 

To explain the additional current $j_3$ , we should note that any squeezed state is characterized by mean excitation number ($<b^{\dagger}b>_s$) and two-excitation correlation function ($<b^2>_s$ and $<b^{\dagger2}>_s$). 
$j_3$ is related to the second property of the squeezed-states.

 Also,  $\Pi_1$  is the part of the entropy production that arises from the temperature difference between the two baths (more exactly, the inverse Bose-Einstein occupations $N_i=2\bar{n}_i+1$). 

The expression for $\Pi_1$ is very similar to what is obtained  in Ref. \cite{Malouf}, for entropy production of a general linear network of harmonic nodes which are coupled to multiple heat baths at different temperatures. In this study, we have an additional term for the entropy production that arises from vacuum fluctuations of the baths. Also, the heat current between the two baths is affected by the parametric amplification strength $\Lambda$. 

We can reexpress $\Pi_1$ as
 \begin{equation}\label{pi_1,J}
 \Pi_1=(J_{12}+J'_{12})(\frac{1}{N_2}-\frac{1}{N_1}),
 \end{equation}
where $J_{12}=G a_{11}(N_1-N_2)$ is the energy current due to the temperature gradient between the baths. The additional current $J'_{12}=-(\Omega-\omega_m)(N_1-N_2)a_{31}$ arises from temperature gradient and frequency gradient $(\Omega-\omega)$. As can be seen in the drift matrix of Eq. (\ref{drift}), due to the parametric amplification process, the quadratures of the bosonic mode 2 oscillate at different frequencies $\omega$ and $\Omega$, and this difference produces  “parametric current"  $J'_{12}$ within the system. 
Since $a_{11}$  and $a_{31}$  are always positive (see Appendix \ref{Appendix:A} for details), the  two currents have  opposite signs. Hence, for $N_1>N_2$ , we have  $J_{12}>0$, which describes the heat flow from the hot thermal bath  to the cold thermal bath. Also, the parametric current $J'_{12}$ is the  current in the opposite direction from the  cold thermal bath  to the hot thermal bath.
Also, since $Ga_{11}>(\Omega-\omega)a_{31}$  (see Appendix \ref{Appendix:A} for details) the total energy current ($J=J_{12}+J'_{12}$) has the same sign as $N_1-N_2$ and $\Pi_1$ is always positive.

It is easy to show that for $\Lambda=0$ and  in the classical regime where $\hbar\omega\ll k_B T_2$ and $\Delta\approx\omega$, $\Pi_1$ takes the following familiar form
\begin{equation}
\Pi_1\approx J_0 (\frac{1}{T_2}-\frac{1}{T_1}),
\end{equation}
which is the Onsager entropy production between two systems kept at different temperatures\cite{Onsager}. The heat current in this regime is given by $J_0 =\frac{\hbar\omega G a_0(N_1-N_2)}{2k_B}$ ( the exact expression for $a_0>0$ is given in  appendix\ref{Appendix:A}). 

It is obvious that for  $N_{i}\gg 1,(i=1,2)$, the  Onsager part  of the entropy production $\Pi_1$ is more important than the vacuum part $\Pi_0$. And for $N_2=N_1$, $\Pi_0$ is the only part of the entropy production.
\begin{figure}[h]
\centering
\includegraphics[width=7.5cm,height=3.5cm,angle=0]{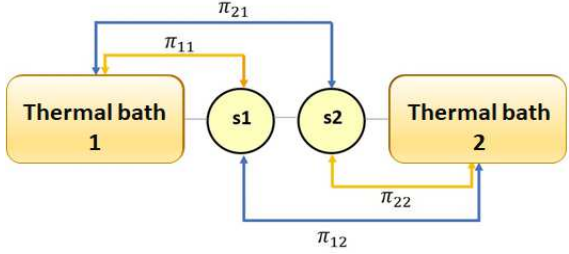}
 \caption{(Color online) Different components of the vacuum entropy production. The entropy production of each subsystem is affected by a local and a non-local thermal bath}.
 \label{Fig2:scheme2}
\end{figure} 

To get more insights into the vacuum entropy production $\Pi_0$, we  decompose $\Pi_0$ into four terms
\begin{equation}\label{pi0-2}
\Pi_0=\pi_{11}+\pi_{22}+\pi_{12}+\pi_{21},
\end{equation} 
where  $\pi_{ii}$ are the parts of the vacuum entropy production due to the coupling between mode $i(i=1,2)$ and its local bath $i$ and  $\pi_{ij\neq i}$  are the parts of the vacuum entropy production due to the coupling between mode $i(i=1,2)$  and its non-local bath $j\neq i$.  These terms are given by
\begin{subequations}\label{pi_m,f}
\begin{eqnarray}
\pi_{12}&=&G a_{22},\quad \pi_{2,1}=Ga_{11}-4\Lambda a_{31},\\
\pi_{11}&=&G a_{21},\quad \pi_{2,2}=Ga_{12}-4\Lambda a_{32},
\end{eqnarray}
\end{subequations}
To explain the naming of these terms, we should go back to  Eqs. (\ref{vii}) and (\ref{co-terms-1}) to express the mean excitation number of each oscillator in terms of input noise correlations. According to these equations, the mean excitation number of each mode in the steady-state $(N_i^s)$ is given by
\begin{subequations}\label{N_is-2}
\begin{eqnarray}
N_1^s&=&\sigma_{11}^s+\sigma_{22}^s\nonumber\\&=& (1+\frac{Ga_{21}}{2\kappa})N_1+\frac{G a_{22}}{2\kappa} N_2 ,\\
N_2^s&=&\sigma_{33}^s+\sigma_{44}^s\nonumber\\&=& (1+\frac{Ga_{12}-4\Lambda a_{32}}{2\gamma})N_2+\frac{Ga_{11}-4\Lambda a_{31}}{2\gamma} N_1,\quad\quad
\end{eqnarray}
\end{subequations}
and hence, the mean excitation number of  mode $i$ is affected by the input-noise correlations of its local thermal bath $(N_i)$  and input-noise correlations of its non-local thermal bath $(N_{j \neq i})$. Therefore, $G a_{21}$ quantifies the effect of local bath 1 on the mean excitation number of mode 1, $G a_{22}$ quantifies the effect of local bath 2 on the mean excitation number of mode 1, and so on (Fig. \ref{Fig2:scheme2}).

It is easy to show that  $Ga_{11}-4\Lambda a_{31}=G a_{22}>0$ and hence, we have $\pi_{12}=\pi_{21}>0$ (see Appendix  \ref{Appendix:A} for more details).
On the other hand, $\pi_{11}$ and $\pi_{22}$ are not equal in general, and depending on the parameter values can be positive, negative, or zero.

Another interesting point that should be noted here is  that for $N_1\neq N_2$,  we have $J_{12}+J'_{12}=\pi_{2,1}(N_1-N_2)$, i.e., the heat current is only related to the coupling between mode $i$ and its non-local thermal bath $j \neq i$ in the steady-state.
 
\section{Numerical Calculation of the Steady-State Vacuum Entropy Production}\label{NUMERIC}
To get additional insights into the analytical results of the previous section, we present some numerical results for $\Pi_0$ and its components in different regimes. 

\begin{figure}
\centering
 \includegraphics[width=4.2cm,height=3.86cm,angle=0]{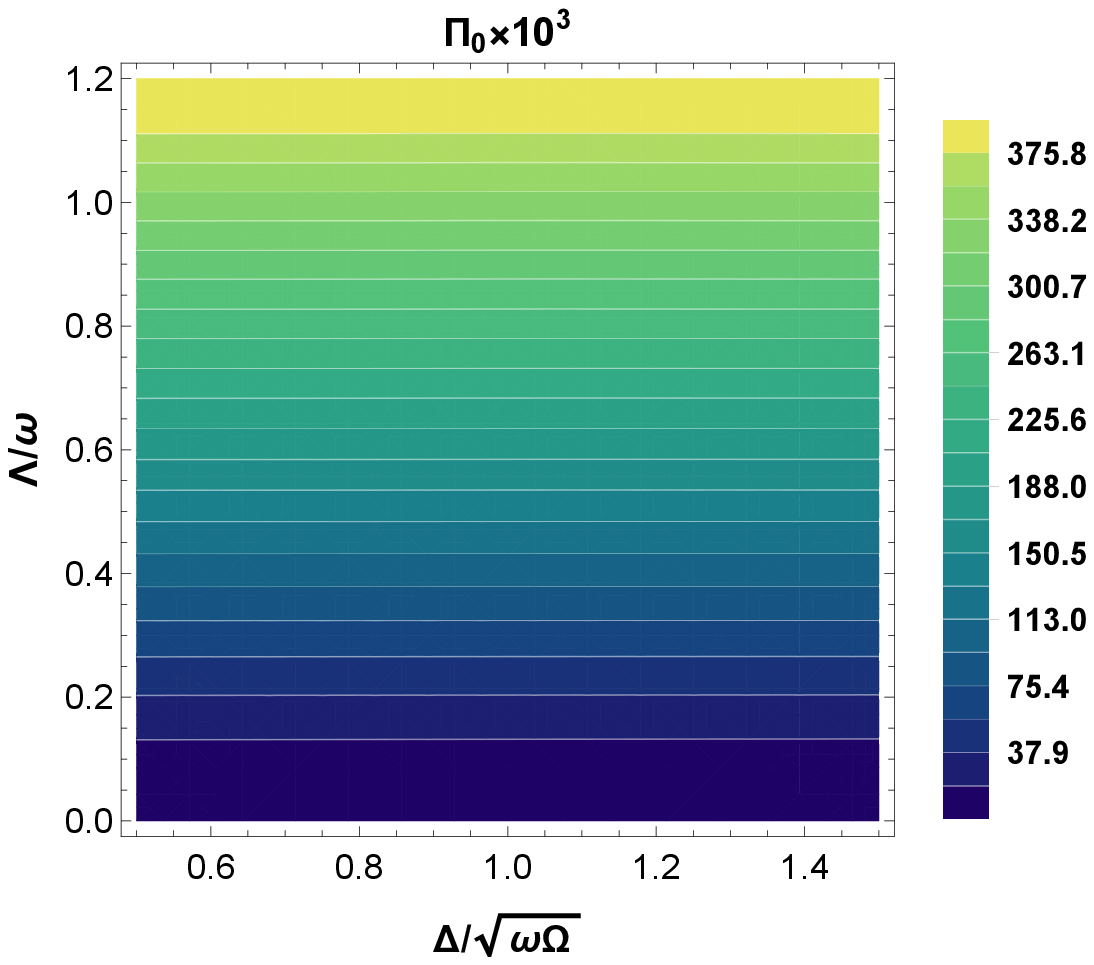}
 \includegraphics[width=4.2cm,height=3.86cm,angle=0]{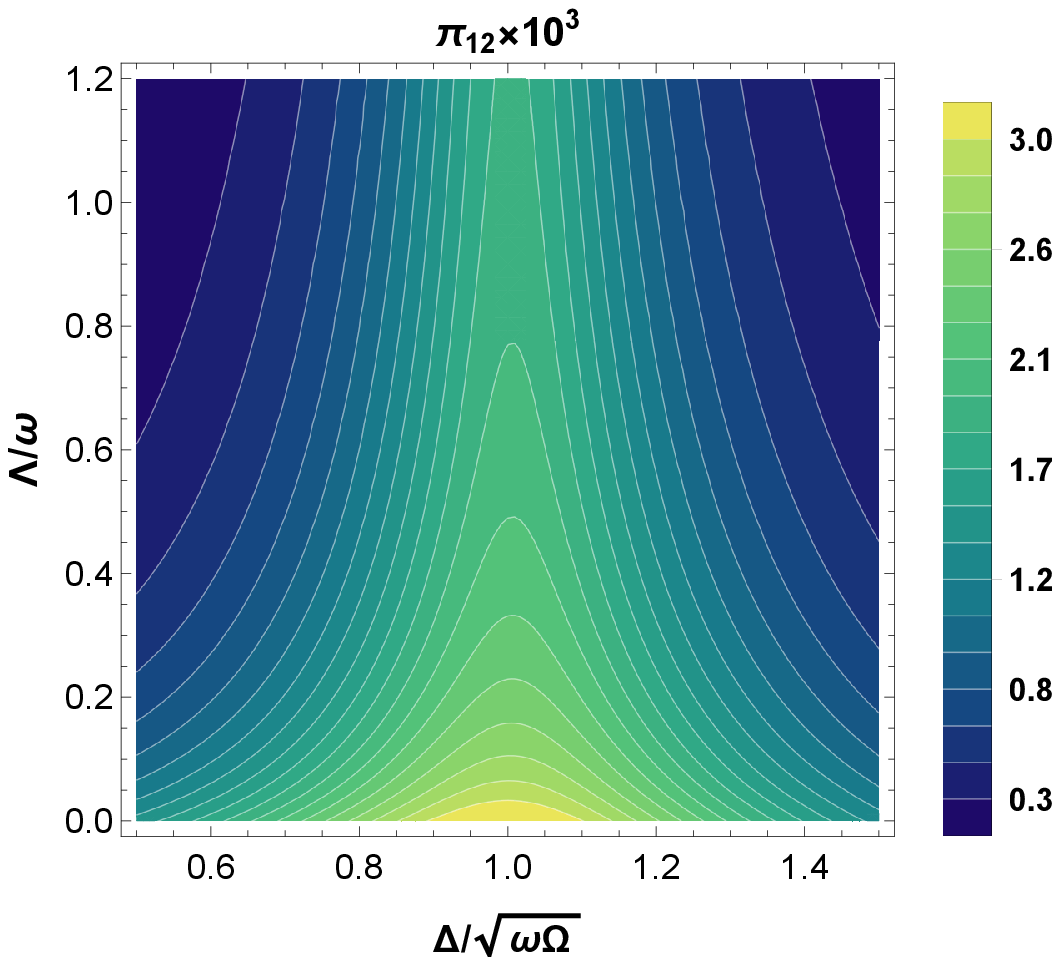}\\
 \includegraphics[width=4.1cm,height=3.86cm,angle=0]{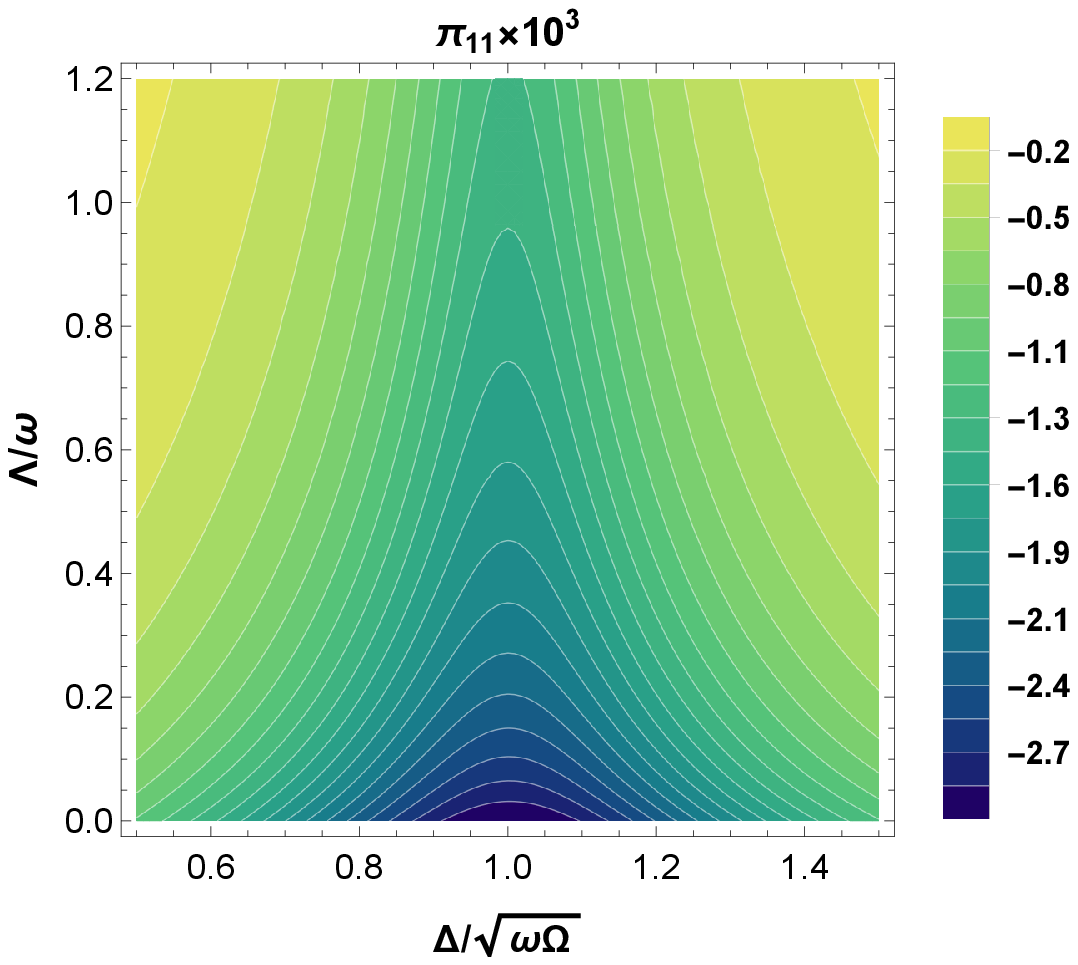}
 \includegraphics[width=4.1cm,height=3.86cm,angle=0]{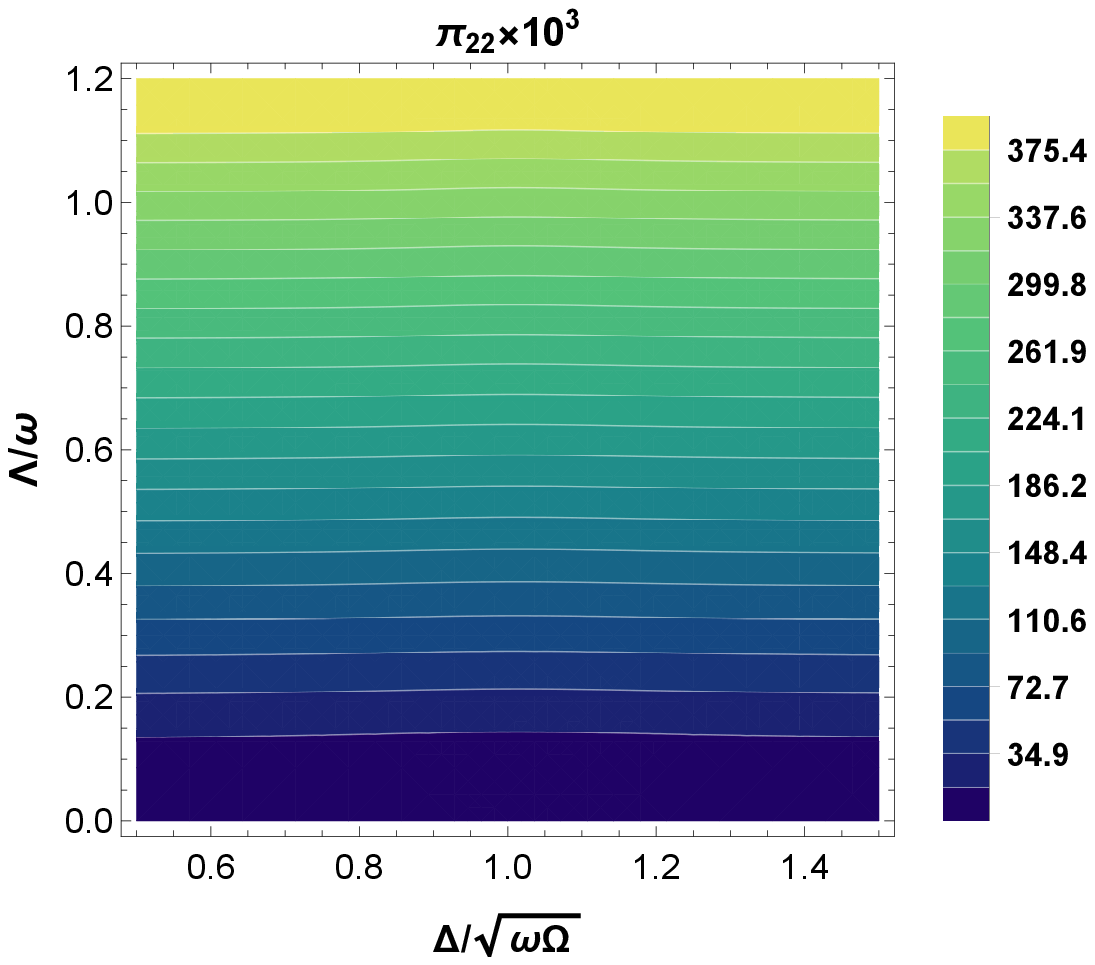}
\caption{(Color online)  Vacuum entropy production rate  and its components for $N_m=N_f$ versus $\Delta/\sqrt{\Omega\omega}$ and $\Lambda/\omega$. The parameters are $\kappa=\gamma=0.2\omega$, $G=0.05\omega$, $\omega=1$. }
\label{Fig:pi0-1}
  \end{figure}

In Fig. \ref{Fig:pi0-1}, $\Pi_0$ and its components are plotted as a function of  $\Delta/\sqrt{\Omega\omega}$ and $\Lambda/\omega$ for $G<\kappa,\omega$. As can be seen, $\pi_{12}$ maximizes at resonance condition $\Delta=\sqrt{\Omega\omega}$. This is the frequency in which the beam splitter-like term of the interaction Hamiltonian $g(a^{\dagger}b+ab^{\dagger})$ is pronounced and the subsystems exchange excitations (Note that this is also the frequency in which the total heat current maximizes). 
 \begin{figure}
\centering
 \includegraphics[width=4.2cm,height=3.87cm,angle=0]{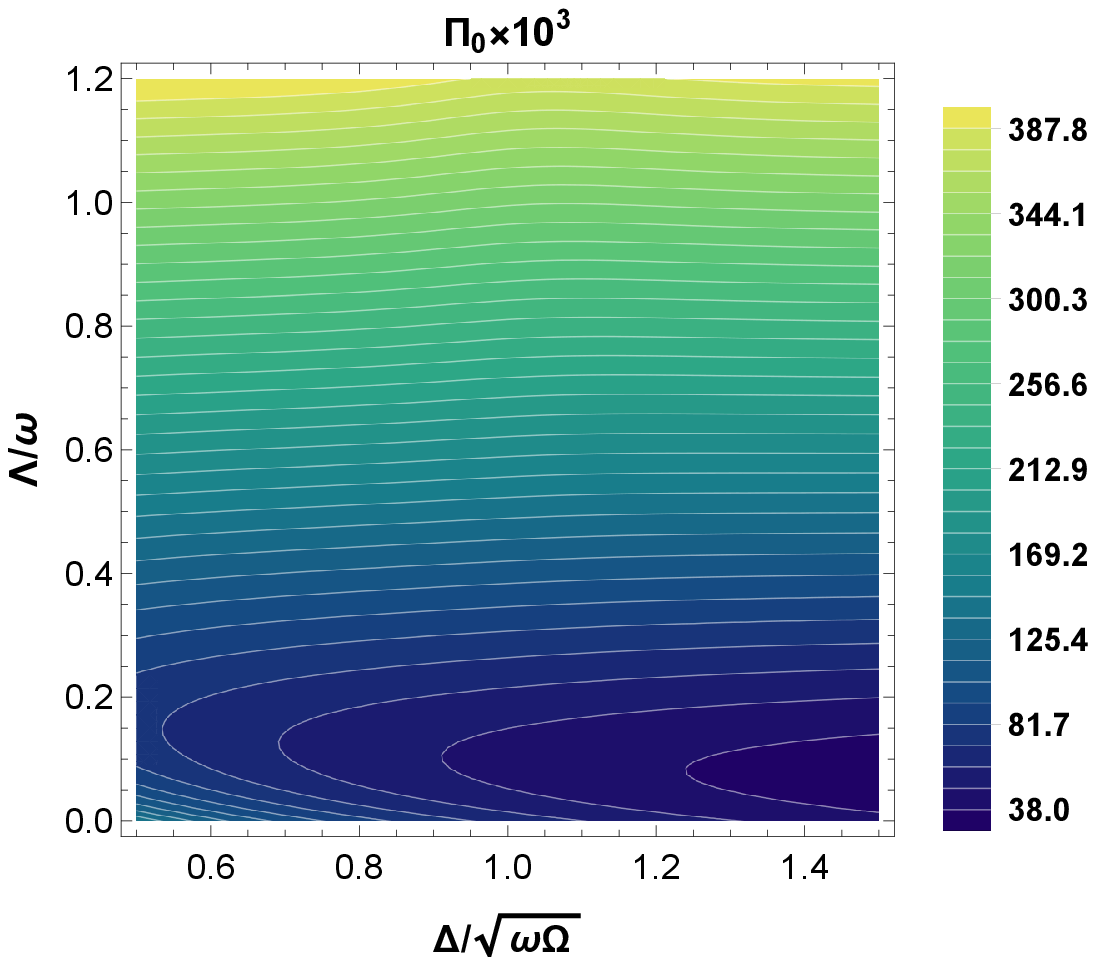}
 \includegraphics[width=4.2cm,height=3.87cm,angle=0]{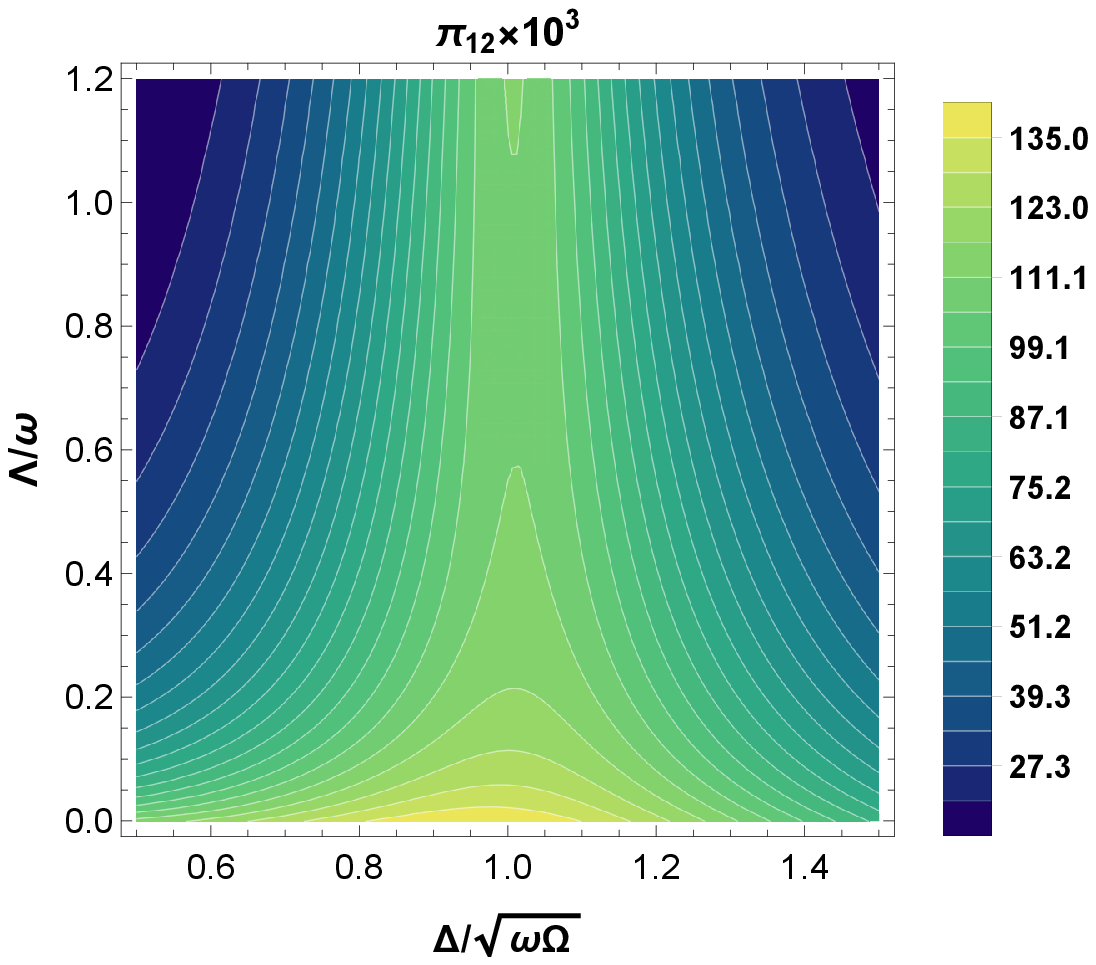}\\
  \includegraphics[width=4.2cm,height=3.87cm,angle=0]{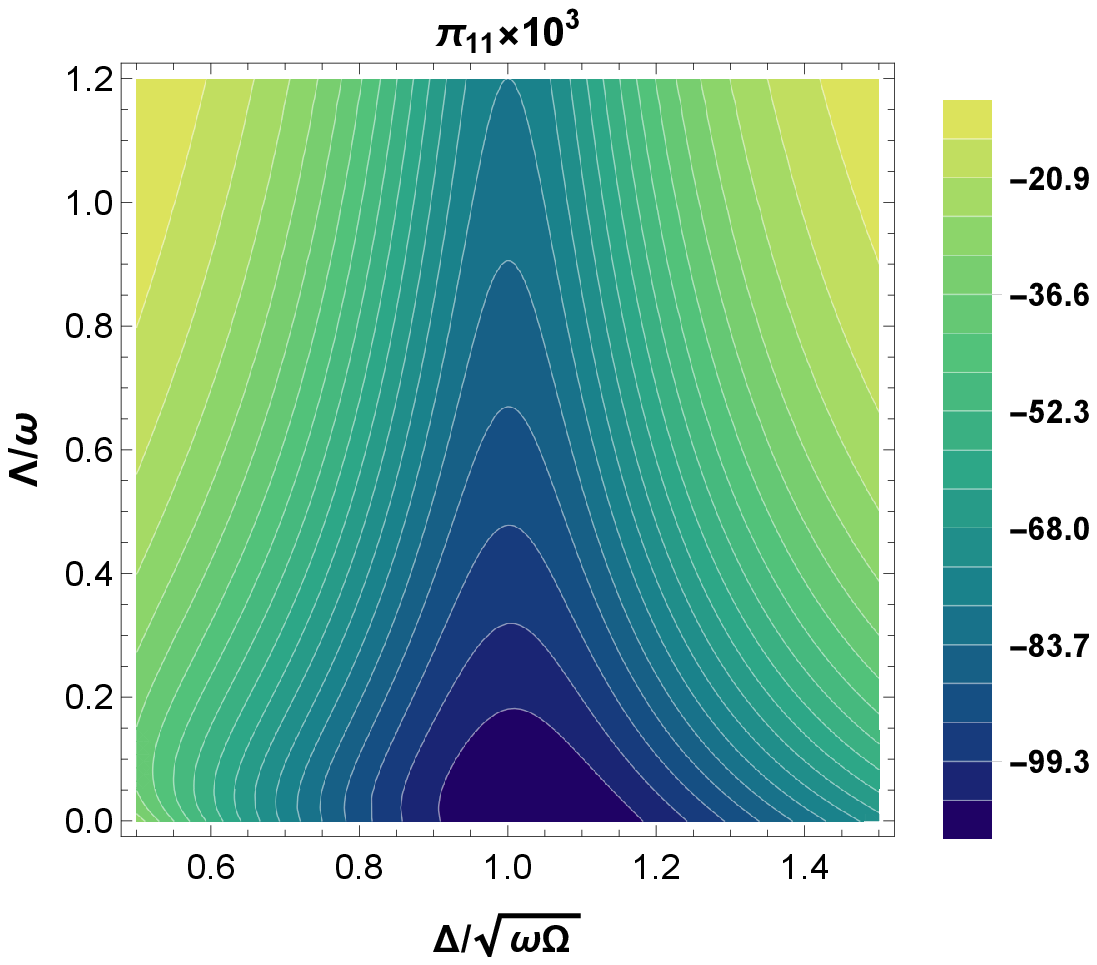}
  \includegraphics[width=4.2cm,height=3.87cm,angle=0]{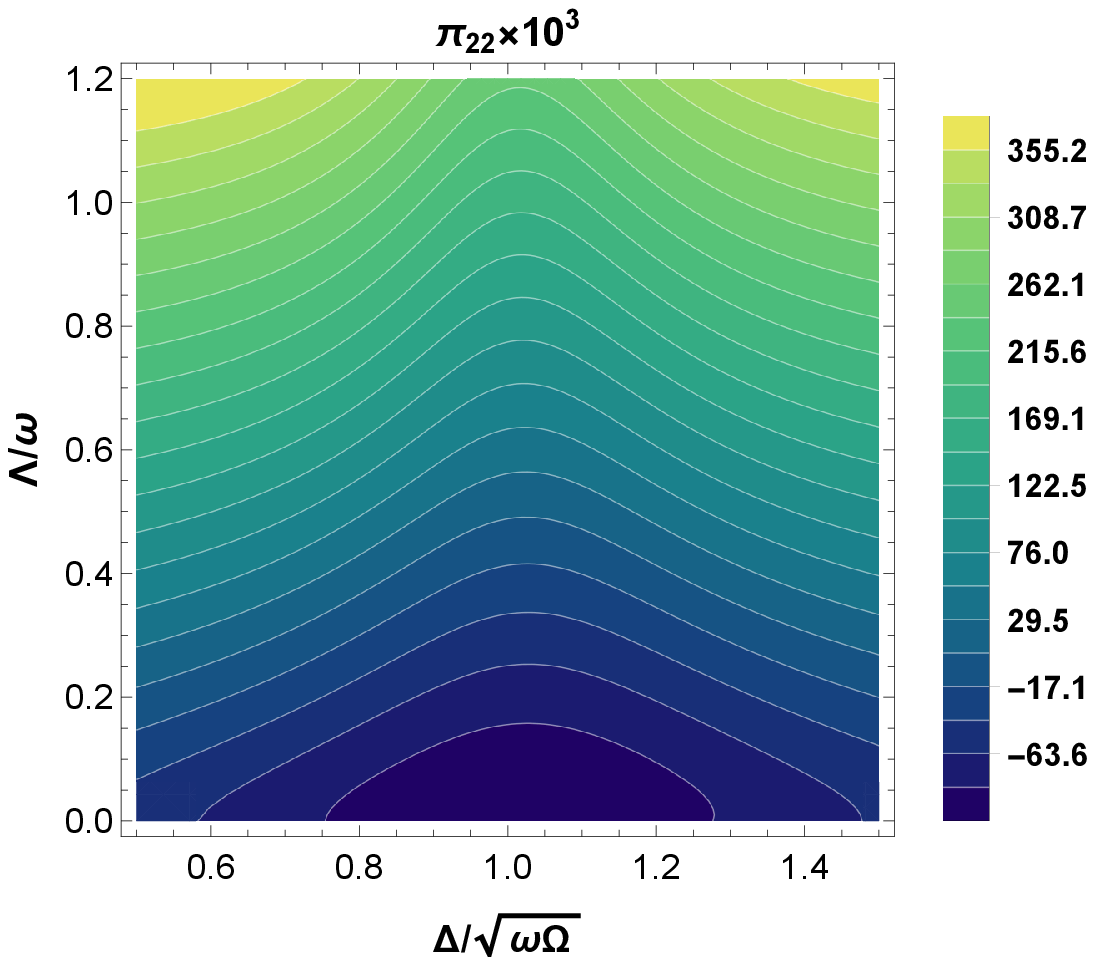}
 \caption{(Color online)  Vacuum entropy production rate  and its components  versus $\Delta/\sqrt{\Omega\omega}$ and $\Lambda/\omega$ close to the resonance frequency. The parameters are $\kappa=\gamma=0.2\omega$, $G=0.5\omega$, $\omega=1$. }
 \label{Fig:pi0-2}
 \end{figure}

In this regime, $\pi_{11}$ has a very small negative contribution to the total vacuum entropy production. However, $\pi_{22}$  has the main positive contribution to the total vacuum entropy production. It is easy to show that in this regime $\pi_{22}$ is approximately given by 
\begin{equation}
\pi_{22}\simeq-4\Lambda a_{32}\simeq\frac{8\gamma\Lambda^2}{\Gamma^2},
\end{equation}
which is always positive. The positivity of $\pi_{22}$ in this range is related to the negativity of the self-correlation term $\sigma_{34}^s\simeq a_{32}/2$ and squeezing of  subsystem 2 in  $q$ (Note that according to Eq. (\ref{Eq:sigmaij}) we have  $\sigma_{33}^{s}<1/2$ in this regime). Therefore, self-correlation between the quadratures of the second mode is responsible for pushing the second mode away from the thermal equilibrium state towards the squeezed state and increasing vacuum entropy production of the total system.

 Figure \ref{Fig:pi0-2} shows  the vacuum entropy production and its components for  $\kappa<G<\omega$ for $\Delta$ close to the resonance frequency. 
It shows that  in the range of parameters in which $\pi_{12}$ is maximum ($\Delta\simeq\sqrt{\Omega\omega}$ and $\Lambda<0.2\omega$),  $\pi_{11}$ and $\pi_{22}$ are negative. This implies that  the contribution of each bath to the vacuum entropy production is twofold in this range of parameters.
First, it pulls its local oscillator towards the equilibrium state (by decreasing $\pi_{ii}$ (i = 1,2)) and pushes the other one away from the thermal equilibrium state (by increasing $\pi_{ij\neq i}$). In other words, due to the coupling between the subsystems, each subsystem is pushed away from thermal equilibrium and its local bath pulls it towards an equilibrium state.
As $\Lambda$ gets larger, $\pi_{12}$ becomes smaller and $\pi_{22}$ becomes larger. Therefore, self-correlation and internal correlation compete with one another at resonance. 

Comparing  the maximum values of $\Pi_0$ in figures \ref{Fig:pi0-1} and \ref{Fig:pi0-2}, it can be seen that for large values of $\Lambda$, the role of coupling strength is very small. 

Also, in both regimes, $\pi_0$ is nonzero for $\Lambda=0$ and $\Delta=\omega$.  In this regime (resonance condition, $\Lambda=0$ and $\gamma=\kappa$), $\pi_0$ and its components are given by
\begin{eqnarray}
\pi_0&=&\frac{G^2\Gamma_1^2\gamma}{d_0},\\
\pi_{12}&=&\frac{G^2\Gamma_1^2\gamma(4d_0+4\gamma^2\Gamma_1^2+G^2\omega^2)}{4(G^2\omega^2+4\gamma^2\Gamma_1^2)d_0},\\
\pi_{11}&=&\pi_{22}=\frac{G^2\Gamma_1^2\gamma\omega^2(5G^2-4\Gamma_1^2)}{4(G^2\omega^2+4\gamma^2\Gamma_1^2)d_0},
\end{eqnarray}
where $d_0=\Gamma_1^4-G^2\omega^2>0$ as a result of the stability condition and  $\Gamma_1=\sqrt{\omega^2+\gamma^2}$. As expected $\pi_{12}$ is always positive and $\pi_{11}(=\pi_{22})$ is always negative. Therefore, the composite system remains in the nonequilibrium state because of the interaction. It is predictable that vacuum entropy production can be suppressed when the interaction Hamiltonian is  subjected to the constraints of thermal operations ($[H_0,H_{int}]\neq 0$)\cite{Oppenheim,Lostaglio1,Horodecki,Lostaglio2,Cwiklinski,Misra, Das}. A detailed discussion of this issue will be given elsewhere.

 \section{Conclusions}\label{Sec:Conc}
We have studied  entropy production rate for a composite Gaussian system. The modes are exposed to the effects of local thermal baths. One of the coupled modes  is driven into the thermal-squeezed state through the parametric amplification process. Our analytical expression for steady-state Wigner entropy production shows that it depends on the internal correlations between the modes and self-correlation between non-commuting operators of the parametrically driven mode.
The self-correlation squeezes the driven mode, produces a heat current from cold to the hot bath, and reduces the total heat current between the reservoirs at different temperatures. 
Also, we have demonstrated that entropy production is nonzero, even when the heat flux between the baths is zero (i.e., when they have identical thermal occupation numbers or when they are in the vacuum state). 
By recognizing the local and non-local effects of the baths on each mode, we have shown that self-correlation enhances the local contribution of one bath and internal correlation increases the nonlocal contributions of both baths to the total entropy production and they compete with one another in the steady-state.
These results imply that, unlike previous claims\cite{Huang,Rosnagel,Abah,Alicki,Niedenzu1,Klaers,Agarwalla,Niedenzu2,Manzano-16,Manzano-sq,Manzano-2021}, squeezing may limit the efficiency of  non-equilibrium powerful heat engines in two ways. It reduces the heat current between the baths and  increases the effect of vacuum fluctuations of the baths on the total entropy production. 

It should be noted that, in the current study, the interaction Hamiltonian is not subjected to the constraints  of thermal operations. It is predictable that considering the constraints  of thermal operation reduces the vacuum entropy production and irreversible flows, but we leave this for future studies.


\section*{Acknowledgements}
S.Sh. is grateful to  A. Rezakhani and S. Alipour  for useful
discussions and comments. M.R. would like to thank M. Huber for his useful comments.



\appendix
\section{}\label{Appendix:A}
As stated, the correlation terms can be compactly cast as 
\begin{subequations}\label{co-terms-a1}
\begin{eqnarray}
\sigma_{14}^s&=&\sigma_{41}^s=(a_{11}N_1+a_{12}N_2)/2,\\
\sigma_{23}^s&=&\sigma_{32}^s=(a_{21}N_1+a_{22}N_2)/2,\\
\sigma_{34}^s&=&\sigma_{43}^s=(a_{31}N_1+a_{32}N_2)/2,
\end{eqnarray} 
\end{subequations}
where the coefficients are given by
\begin{subequations}\label{ai,f,m}
\begin{eqnarray}
a_{11}&=&\frac{G\kappa\gamma}{d_1}(\delta^2\Gamma^2 u_1+\eta(\Gamma^2\kappa_1+\delta^2\kappa_2))
,\quad\quad\\
a_{12}&=&\frac{G\Delta\gamma}{\omega d_1} (\delta^2\Gamma_1^2\gamma u_1 -\eta(\kappa\kappa_1(\omega^2+\omega\Omega)\nonumber\\&+&\gamma(u+\kappa_1 \omega^2)
)),\\
a_{21}&=&\frac{G\kappa\omega}{\Delta d_1}(\delta^2\Gamma^2\kappa u_1-\eta (\delta^2 \kappa_2^2+\kappa\gamma \Gamma^2)),\quad\quad\quad\\
a_{22}&=&a_{11}-\frac{G\kappa\gamma}{d_1}(\omega\Omega-\omega^2)(\delta^2 u_1+\eta(\gamma+\kappa)),\quad\quad\\
a_{31}&=&\frac{G^2\kappa\gamma}{ d_1}(\delta^2 u_1+\eta(\gamma+\kappa))\omega,\\
a_{32}&=&\frac{\gamma}{\omega d_1}(\delta^4\Gamma_1^2\gamma u_1+\eta^2(\gamma+\kappa)\kappa_1\nonumber\\&-&\eta\gamma \delta^2(2u+\kappa_1\omega^2-\kappa(\delta^2+\gamma\kappa_1))\nonumber\\&-&
\eta\gamma\kappa(\omega\Omega-\omega^2)(\omega\Omega+\kappa_1^2)\nonumber\\&-&\eta\omega\Omega\delta^2(\gamma^2+\kappa_1\kappa)),
\end{eqnarray}
\end{subequations}
and we have defined
\begin{subequations}
\begin{eqnarray}
&&d_1=2 \eta (-\eta(\gamma + \kappa)^2  + u_1 (\gamma \delta^2 + \kappa 
\Gamma^2)),\quad\quad\quad\\
&&u= \gamma^3 + 4 \gamma^2 \kappa+ \delta^2 \kappa + 
   4 \gamma \kappa^2,\\
  && u_1=u+\gamma\omega\Omega,\\
  && \delta=\sqrt{\Delta^2+\kappa^2},\\
 &&  \Gamma=\sqrt{\gamma^2+\omega\Omega},\\
 &&   \Gamma_1=\sqrt{\gamma^2+\omega^2},\\
 && \kappa_1= (\gamma+2\kappa),\\
 && \kappa_2= (2\gamma+\kappa).
\end{eqnarray}
\end{subequations}

Using the definition of $\eta,$ we can express $d_1$ as
\begin{eqnarray}
d_1&=&2\eta \lbrace\kappa\gamma ((\gamma+\kappa)^4 +2 (\gamma + \kappa)^2 (\Delta^2 + 
\omega\Omega) \nonumber\\&&+ (\Delta^2 -\omega\Omega)^2)
+G^2\Delta\omega (\gamma + \kappa)^2\rbrace,\quad\quad
\end{eqnarray}
which is obviously positive (note that $\eta$ is stability parameter and is positive). 
Also, according to Eqs.(\ref{ai,f,m}a-\ref{ai,f,m}f), $a_{11}$ and $a_{31}$ are  always positive and  we have 
\begin{equation}\label{App1}
Ga_{22}=Ga_{11}-(\Omega-\omega)a_{31},
\end{equation}
Hence, for $\Lambda=0$, we have $a_{22}=a_{11}=a_0$, where
\begin{equation}\label{a0}
a_0=\frac{G\kappa\gamma_m(\delta^2\Gamma_1^2 u_1+\eta_1(\Gamma_1^2\kappa_1+\delta^2\kappa_2))}{2 \eta_1 (-\eta_1(\gamma + \kappa)^2  + u_1 (\gamma \delta^2 + \kappa 
\Gamma_1^2))},
\end{equation}
where $\eta_1=\delta^2\Gamma_1^2-G^2\Delta\omega$ is the stability parameter for the system when $\Lambda=0$.

For $\Lambda\neq0$, the relation between correlation terms in Eq. (\ref{App1}), leads to Eq. (\ref{pi_1,J}) for $\Pi_1$ where the sum of the currents which is given by 
\begin{eqnarray}
J_{12}+J'_{12}&=&\frac{G^2\kappa\gamma_m}{d_1}(N_1- N_2)(\delta^2\Gamma_1^2u_1\nonumber\\&&+\eta(\Gamma^2\kappa+\delta^2\kappa_2+\Gamma_1^2(\kappa+\gamma))).
\end{eqnarray}
has the same sign as $N_1 - N_2$.

According to  Eq. (\ref{pi_m,f}), the elements of $\Pi_0$ are given by
\begin{subequations}
\begin{eqnarray}
\pi_{12}&=&G a_{22},\quad \pi_{2,1}=Ga_{11}-4\Lambda a_{31},\\
\pi_{11}&=&G a_{21},\quad \pi_{2,2}=Ga_{12}-4\Lambda a_{32},
\end{eqnarray}
\end{subequations}
Using  Eq. (\ref{ai,f,m}) for $a_{ij}$, we can reexpress the elements of  vacuum entropy production as  
\begin{eqnarray}
\pi_{12}&=&\frac{G^2\kappa\gamma}{d_1}(\Gamma_1^2\delta^2 u_1+
\eta\kappa(\Gamma^2+\Gamma_1^2+\delta^2)\nonumber\\&+&\eta\gamma(2\delta^2+\Gamma_1^2)),\\
\pi_{11}&=&\frac{G^2\kappa\omega}{\Delta d_1}(\delta^2\Gamma^2\kappa u_1-\eta (\delta^2 \kappa_2^2+\kappa\gamma \Gamma^2)),\\
\pi_{22}&=&\Pi^{(G=0)}+G a_{12}+\frac{\gamma(\Omega-\omega)\Gamma_1^2}{2\Gamma^2\omega}\nonumber\\&-&\frac{\gamma(\Omega-\omega)}{d_1\omega}y_1,
\end{eqnarray}
where
\begin{eqnarray}
y_1&=&(u_1\gamma\Gamma_1^2\delta^4+\eta^2\kappa(\gamma+\kappa)
+\eta\kappa\gamma\Gamma_1^2(\kappa_1^2+\Omega\omega)\nonumber\\
&-&\eta\delta^2(\Gamma_1^2\gamma\kappa_1+\Gamma^2\kappa(\gamma+\kappa))),\quad\quad
\end{eqnarray}
and $\pi_{12}=\pi_{21}$ as a result of Eq. (\ref{App1}).

\end{document}